\documentclass[12pt,preprint]{aastex}
\usepackage{emulateapj5,apjfonts}
\usepackage{graphics}
\usepackage{amssymb}
\usepackage{bbm}
\usepackage{amsmath,textcomp,array}
\usepackage{onecolfloat}
\usepackage{bm}

\lefthead{Heitmann et al.}
\righthead{The Outer Rim Simulation}

\begin{document}


\def\head{

\title{The Outer Rim Simulation: A Path to Many-Core Supercomputers}
\author
{Katrin~Heitmann\altaffilmark{1}, 
Hal Finkel\altaffilmark{2}, Adrian Pope\altaffilmark{3}, Vitali
Morozov\altaffilmark{2}, Nicholas Frontiere\altaffilmark{1,4},  
Salman~Habib\altaffilmark{1,3}, Esteban Rangel\altaffilmark{2}, Thomas
Uram\altaffilmark{2},  
Danila Korytov\altaffilmark{1,4}, 
 Hillary Child\altaffilmark{1,4}, Samuel Flender\altaffilmark{2}, Joe
 Insley\altaffilmark{2}, Silvio Rizzi\altaffilmark{2}} 

\affil{$^1$ HEP Division,  Argonne National Laboratory, Lemont, IL 60439}
\affil{$^2$ ALCF Division,  Argonne National Laboratory, Lemont, IL 60439}
\affil{$^3$ CPS Division,  Argonne National Laboratory, Lemont, IL 60439}
\affil{$^4$ Department of Physics, University of Chicago, Chicago, IL 60637}

\date{today}

\begin{abstract}

  We describe the Outer Rim cosmological simulation, one of the
  largest high-resolution N-body simulations performed to date,
  aimed at promoting science to be carried out with large-scale
  structure surveys. The simulation covers a volume of (4.225Gpc)$^3$
  and evolves more than one trillion particles. It was executed on
  Mira, a BlueGene/Q system at the Argonne Leadership Computing
  Facility. We discuss some of the computational challenges posed by a
  system like Mira, a many-core supercomputer, and how the simulation
  code, HACC, has been designed to overcome these challenges. We have
  carried out a large range of analyses on the simulation data and we
  report on the results as well as the data products that have been
  generated. The full data set generated by the simulation totals more
  than 5PB of data, making data curation and data handling a large
  challenge in of itself. The simulation results have been used to
  generate synthetic catalogs for large-scale structure surveys,
  including DESI and eBOSS, as well as CMB experiments. 
  A detailed catalog for the LSST DESC data
  challenges has been created as well. We publicly
  release some of the Outer Rim halo catalogs, downsampled particle 
  information, and lightcone data.

\end{abstract}

\keywords{methods: N-body ---
          cosmology: large-scale structure of the universe}}

\twocolumn[\head]

\section{Introduction}

\begin{figure}
\centerline{
 \includegraphics[width=3.5in]{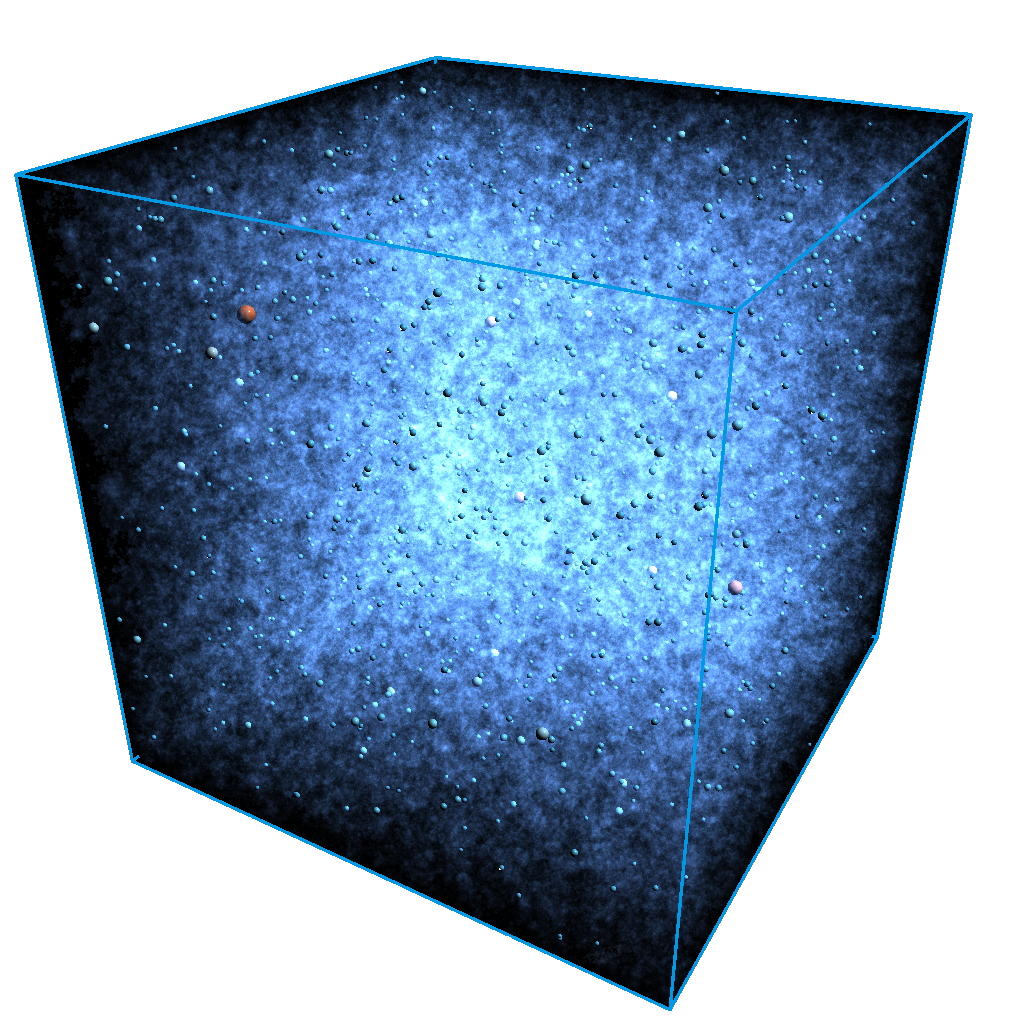}}
\caption{\label{fig:cube}Visualization of the halos in the Outer Rim
  simulation at redshift $z=0$. Halos above a mass of $1.8\cdot
  10^{15}$M$_\odot$ are shown as spheres, capturing $\sim$1000 of the
  heaviest halos, and colored by mass (red indicates more massive),
  while halos down to a mass of $\sim 5\cdot 10^{10}$M$_\odot$ are
  represented by blue Gaussian density ``splats''.} 
\end{figure}

Future large-area cosmological surveys, to be carried out with, e.g.,
the Large Synoptic Survey Telescope (LSST) \citep{lsst,desc}, the Dark
Energy Spectroscopic Instrument (DESI) \citep{desi}, or the Wide Field
Infrared Survey Telescope (WFIRST) \citep{wfirst,wfirst2}, are becoming ever
more complex as telescopes reach deeper into space, mapping out the
distributions of galaxies at farther and farther distances. These
observations provide a treasure trove of information about the make-up
of the Universe and its evolution from the very first moments to
today. Interpreting the observations and extracting knowledge from the
surveys requires sophisticated simulations that track the detailed
formation of structure over time.

These cosmological simulations need to cover large volumes and, at the
same time, provide enough mass resolution to resolve small halos that
host dim galaxies at early epochs. They enable a range of important
tasks: investigations of systematic errors and possible mitigation
strategies, developing and testing complex analysis pipelines and
workflows, providing crucial information for survey optimization,
exploring new probes and cross-correlations among different wavebands,
or forecasts for future survey ideas. In addition, the simulations
play important roles in furthering our understanding of the basic
physics of structure formation.

In order to provide a simulation that can satisfy a substantial range of these
needs, we ran the Outer Rim simulation, one of the largest
N-body gravity-only simulations at the achieved resolution ever carried out.
The Outer Rim simulation covers a volume of (4.225Gpc)$^3$ and evolves
10,240$^3$ particles, leading to a mass resolution of $\sim 2.6\cdot
10^9$M$_\odot$. We saved and analyzed almost 100 time snapshots,
yielding a data volume of more than 5PB. The data products from the
simulation include halo catalogs for different mass definitions,
subhalo catalogs, detailed merger trees, two-point statistics, lightcone 
representations of the data, and subsamples of raw and halo
particles. The high mass and temporal resolution of the simulation and
the corresponding outputs make this data set ideal for survey related
investigations as well as large-scale structure studies. Synthetic
catalogs can be created using a variety of methods, from Halo
Occupation Distribution (HOD) approaches to more detailed
semi-analytic models. The Outer Rim run continues in the tradition of
the Millennium simulation by~\cite{springel05}, with a similar mass resolution but
with a volume coverage increase by more than a factor of 200. This is
essential for capturing galaxy clustering at large length scales and
for achieving the needed statistics for cluster cosmology.

The simulation was carried out with the Hardware/Hybrid Accelerated
Cosmology Code (HACC) described in great detail in~\cite{habib14}. The Outer
Rim simulation used a version of HACC that has been specifically
optimized for high-performance on the BlueGene/Q (BG/Q) system Mira,
using a tree implementation for the short-range solver as well as
individual particle time stepping once the clustered regime is
reached. Due to the vast size of the simulation, some of the analysis
tasks also posed new challenges as described in this
paper. Figure~\ref{fig:cube} shows an example visualization of the
halos extracted from the simulation.

In this paper we describe the Outer Rim simulation and related data products in
detail, some of which will be publicly released. We
also discuss some of the computational implementations on the BG/Q
that enabled us to carry out the simulation efficiently and how these
developments also apply to next-generation supercomputers. Next
we show a set of selected results obtained from the simulation so
far. Some of these results have been used already by eBOSS~(\citealt{eboss1,eboss2,eboss3}) and
in a detailed study of the halo concentration-mass relation by~\cite{child18};
others will be discussed in more detail in accompanying papers in the
near future.

The paper is organized as follows. First, in Section~\ref{hacc}, we
give a brief description of HACC, focusing on the BG/Q and therefore
on the many-core optimizations employed. We also provide a short
summary of how we combat the analysis workflow challenges and the I/O
performance achieved on Mira. In Section~\ref{OR_sim} we describe the
data products from the simulation and highlight some of results
obtained so far. More detailed analyses of the data covering a wide
range of scientific questions are in preparation; this first paper
will serve partially as a motivating reference for these upcoming
studies. In Section~\ref{sec:release} we briefly describe the
data we release from the Outer Rim. The data release is part of a larger project described
in an accompanying paper. We conclude and provide a brief outlook in
Section~\ref{conclusion}. 

\section{HACC}
\label{hacc}

The HACC code has been developed over the last several
years~\citep{habib09,habib12,habib13,habib14} with the specific aim of
providing a code that runs at high performance on a variety of
architectures without having to implement intrusive hardware-dependent
changes. As often used in cosmological N-body codes, the total force
evaluation in HACC is divided into an FFT-based long-range force
solver and a short-range force solver. In the current version of HACC,
most recently described in~\cite{habib14}, 95\% of the code base (the
long-range force solver and the MPI-based communication layer) remains
unchanged when moving between different platforms while the remaining
5\% (the on-node implementation of the short-range solver) is
optimized given the specification of the hardware. This optimization
includes algorithmic changes as well -- depending on costs of
computation vs. data movement, in some cases tree-based algorithms are
optimal, while in other cases direct particle-particle interactions may
be preferable. Given that most of the code stays unchanged, we refer
to this approach as ``soft-portability'', compared to a hard-portable
code that stays completely unchanged and the compilers on the
different systems take care of all the optimization. Currently, HACC
itself, in this sense, is soft-portable, while the analysis set up is
hard-portable, relying on the NVIDIA Thrust library.

Originally developed for the Cell-accelerated machine
Roadrunner~(\citealt{habib09,hacc2}) -- keeping soft-portability in mind even at that
early stage -- HACC currently runs on a diverse range of
architectures, including X86, BG/Q, GPU-accelerated systems, and most
recently Intel's Knights Landing (KNL), which powers the most recent
supercomputers installed at ALCF and NERSC. \cite{habib12} provide
some details about the original BG/Q implementation and optimization
while \cite{habib13} focus on the portability aspect of HACC and
demonstrate outstanding performance on Titan, a GPU accelerated
supercomputer. In the following we focus on the BG/Q specific aspects
of the HACC short-range solver, given that the Outer Rim simulation
was carried out on Mira, the BG/Q system at the ALCF. For a detailed
discussion of the implementation of the long-range solver, the reader
is referred to~\cite{habib13}. We also comment on how our effort on
the BG/Q translates into a rather straightforward port to KNL-based
systems.

\subsection{HACC on Many-Core Systems}

In this section, we focus on the on-node short-range solver for
many-core systems implemented in HACC. The short-range solver in this
case uses a tree-based algorithm. In order to further increase
parallelism, and to decrease the amount of pointer-chasing relative to
numerical computation during the short-range force calculation, we
implemented a scheme which builds multiple
recursive-coordinate-bisection (RCB) trees (or ``bushes'') on
spatially-disjoint regions instead of one large tree covering
the entire volume of one rank. A physical scale is chosen, between 1-2$h^{-1}$Mpc
for the Outer Rim simulation, depending on the clustering state,
\begin{figure}[b]
  \centerline{
 \includegraphics[width=2.5in]{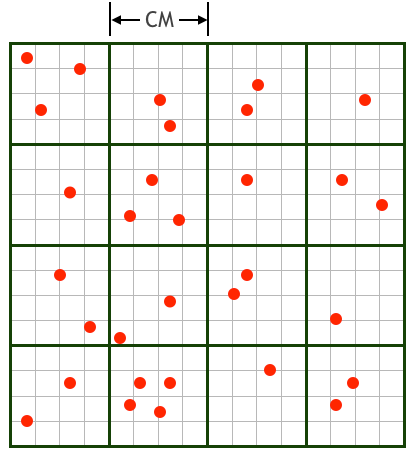}}
\caption{\label{fig:cm2d} Simplified schematic view of a single-rank
  tree structure shown in two dimensions. HACC has two parameters that
  can be tuned to optimize time-to-solution on many-core systems, the
  chaining mesh (CM) size and the final number of particles on a leaf
  at which the tree bisection stops. }
\end{figure}
and the particles are first sorted into bins associated with chaining
mesh (CM) cells on a regular spatial grid. Each cell, boundary cells
excluded, are cubic with a linear size of the chosen scale. Within
each cell, an RCB tree is built. Figure~\ref{fig:cm2d} shows a
simplified schematic view of this scenario in two dimensions. The
chaining mesh cell size and the number of particles per leaf node at
which to stop the tree bisection are input parameters to the code.
These two parameters provide significant flexibility in tuning the
short-range solver on different architectures for achieving the
fastest time-to-solution. In general, higher clustering at late times
is handled better with bigger chaining mesh sizes, while less
clustered distributions require a smaller chaining mesh size. We
accordingly adjust the CM scale during the course of the
simulation. The chaining mesh building process is trivially
parallelizable because the cells are spatially disjoint, and this
parallelization is important to the overall thread scaling of the
code. The individual trees are also less deep, especially those
without many particles in the clustered regime, which might be
trivial, providing a small practical performance advantage.
This is true even though, for each leaf node in each RCB tree, forces
must be accumulated from 27 trees (the parent tree itself plus 26
neighbors). Each tuple of <leaf node, neighboring tree> is added to a
work queue, and this queue is processed by all threads using a
dynamically-scheduled OpenMP parallel loop. Because multiple threads
might end up trying to update forces on particles in the same leaf
node concurrently, locks are used to protect the updates. For the
Outer Rim simulation, one lock per tree was used. 

The transition from the BG/Q architecture to the KNL systems was
straightforward. One lock per leaf node was used to decrease lock
contention and increase thread scalability. Prior to work on KNL
systems, a version of the pair-wise short-range particle interaction
kernel had been implemented for the AVX2 instruction set using
intrinsics, relying heavily on the experience gained from developing
the QPX intrinsics kernel for the BG/Q architecture. For KNL systems,
the AVX2 intrinsics were replaced with the corresponding AVX512
intrinsics, with some minor changes to take advantage of an improved
inverse square-root instruction available on KNL, and the optimal
inner loop unrolling factor was determined experimentally. Finally, a
number of tests were run to optimize the balance of MPI ranks and
OpenMP threads per node and to tune the parameters that determine the
spatial extent and depth of the RCB trees.

\subsection{I/O Performance on the Mira GPFS File System}

In order to obtain optimal I/O performance on a range of large-scale
parallel file systems, we developed a custom I/O implementation for
HACC, called GenericIO. GenericIO is a write-optimized library for
writing self-describing scientific data files and is publicly
available\footnote{http://trac.alcf.anl.gov/projects/genericio}. Mira
is connected to an IBM General Parallel File System (GPFS). Optimal
I/O timing for the raw particles was obtained when writing out the
data into 256 files. The write-speed was approximately 0.15TB/s which
allowed us to write raw particle files in just under 5 minutes. In our
current implementation, the number of analysis files written during
in-situ analysis is locked to the same number of files as the raw
outputs. At early steps this can lead to a relatively large number of
small files, e.g., the number of halos is tiny at very high redshifts
and most files therefore will be empty. This approach then leads to
rather slow I/O and we switch from Posix-I/O to collective MPI-IO by providing a
switch coupled to the size of the output. For outputs smaller than a
certain size, MPI-I/O is used while for large files Posix-I/O is
chosen. The file size above which to use Posix-I/O is an easily
adjustable parameter in the HACC input deck. A more detailed
description of GenericIO including comparison to the performance of
PNetCDF is given in~\cite{habib14}.

\subsection{Analysis Workflow}

\begin{figure}[t]
  \centerline{
 \includegraphics[width=3.5in]{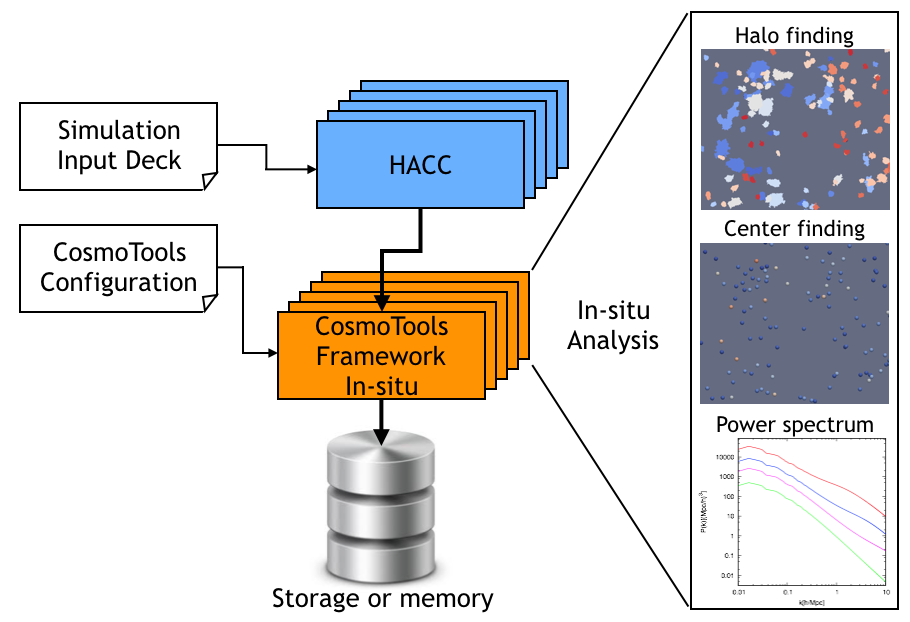}}
\caption{\label{fig:workflow} Schematic description of our current
  workflow implementation in HACC if the analysis is run in-situ. In
  addition, raw HACC data can be written directly to storage and
  CosmoTools can be run on it later. When carrying out the Outer Rim
  simulation, CosmoTools was in an early stage of development and has
  been considerably enhanced since then. }
\end{figure}

A simulation of the size described in this paper poses an enormous
challenge with regard to analysis tasks, in particular where the raw
particle data is concerned. Due to memory demands, handling the raw
data usually requires as much of the resource as running the
simulation itself (in the specific case described in this paper, 32
racks of Mira, out of a total of 48) and therefore is extremely
expensive. Just reading in the data for analysis even with a highly
optimized I/O approach that we developed for Mira, costs close to 200k
core hours for a run of the size of the Outer Rim simulation,
equivalent to running a medium-size simulation. Therefore, if
possible, it is important to avoid reading in the raw data and instead
to carry out analysis tasks on the fly while the raw data is still available
in memory. This in turn requires writing fast, load-balanced, and
memory efficient analysis tools. As part of HACC we have developed and
continue to enhance CosmoTools, a library of high-performance analysis
tools that can be run in-situ or off-line (for a description of an
earlier version, see \citealt{habib14}). As shown in
Figure~\ref{fig:workflow}, CosmoTools is triggered from the simulation
input deck by a simple on/off switch and the specification of the
analysis input parameter file, the CosmoTools configuration file. This
configuration file contains a list of the tools to be run, the list of
snapshots when they are run, and the tool specific parameters, such as
linking length, minimal halos mass, or mass definition for the halo
finder. 
It is also possible to exclude analysis tools when compiling and
building HACC. This option is included to allow the use of external
libraries within analysis tools. The advantage of external libraries
is that they are often highly tuned to deliver good performance, the
disadvantage is that they might not be available on the platform the
simulation is run on. In order to guard for this possible problem,
analysis tools can be excluded during the build process. (An example
for this is a density estimator that depends on the Qhull library.)

Throughout the simulation, we periodically (in-situ) evaluate the
matter power spectrum; this calculation is rather inexpensive but very
helpful to monitor the health of a run.  The halo finder is also
frequently run in-situ. As we describe in the next section, halo
finding is carried out including a range of sub-analysis tasks beyond
the halo finding such as detailed tracking of halo cores over time and
saving all particle identifiers (``IDs'') of particles that reside at
some point in halos. Storing all this information allows us to build
many more data products off-line without requiring very large
resources since the halo finding step reduces the data amount
considerably. It also ensures enough flexibility to carry out analyses
on the data products exploring different parameter settings. This
opportunity is very important since with a simulation like the Outer
Rim run, we entered new regimes for a number of analysis tasks.

In the case of high mass resolution simulations at late times, when
large clusters have formed, some of the halo analysis tasks, like
evaluating the halo gravitational potential minimum for center
finding, can become rather expensive. As detailed in~\cite{sewell15}
we therefore implemented a workflow option where only the small and
medium size halos are analyzed in-situ while all halo particles from
the very large halos are written out to disk and are then analyzed in
post-processing. Usually, only a very small percentage of the halos
will be analyzed off-line this way (so storage requirements are
minimal) but the approach saves a substantial amount of computer
time. \cite{sewell15} provide detailed timing information based on a
large simulation and show different analysis workflows to optimize the
the usage of core hours.

For the Outer Rim simulation we started the halo finder in-situ but
encountered some memory bottlenecks when reaching lower redshifts,
demonstrating that an extreme simulation like this one always uncovers
new challenges that were not apparent when carrying out medium-scale
simulations. Once the stage was reached where cluster sized halos are
forming we switched to off-line analysis in order to further develop
the approach described above where we divide halos into two mass
classes and analyze them separately. In the meantime we also improved
the memory efficiency of the halo finder considerably to enable the
analysis of the final time snapshots.

\section{The Outer Rim Simulation}
\label{OR_sim}
In this section we provide details about the simulation, including a
list of saved outputs and common statistics measured at various
redshifts.  The simulation used 32 racks of Mira, the BG/Q
supercomputer at the Argonne Leadership Computing Facility, each rack
hosting 1024 nodes, each node consisting of 16 user-accessible
cores. During the earlier phase of the simulation, we used 8 ranks per
node and 8 threads, while we switched to 1 rank per node and 64
threads during the later, highly clustered stages of the simulation,
approximately at $z\sim 0.6$. At this point we also switched on the
individual particle time stepper instead of evolving each particle
with the same number of time steps.

The overall size of the dataset generated exceeds 5PB, providing an
unprecedented amount of detailed information about the formation of
the structures in the simulation over time. We classify our outputs
into three levels. Level 1 data is the raw particle data or density
and potential fields that cover the full volume as well as summary
statistics obtained directly from the raw data, such as the matter
power spectrum. Level 2 data is obtained by carrying out an analysis
step on Level 1 data to generate, e.g., halo catalogs and summary
statistics describing the Level 2 data, such as mass functions. Level
2 data is usually at least an order of magnitude smaller than Level 1
data. Finally, Level 3 data is derived from Level 2 data and often
already closer to actual observables, such as galaxy catalogs or sky
maps at different wavelengths. In the following we will specify what
level the data products belong to explicitly.

\subsection{Parameters}
The cosmology used for the simulation is close to the best-fit model
determined by WMAP-7~\citep{wmap7}. This is the same cosmology
underlying two related simulations, the Q Continuum simulation,
described in~\cite{QCont} and the MiraU simulation. The MiraU
simulation was used for example in~\cite{flender} to investigate and
model the kinematic Sunyaev-Zel'dovich effect and in~\cite{MiraU} to
study requirements for precision predictions for next-generation dark
energy surveys. This set of simulations covers a range of mass
resolutions, from $\sim 10^8$M$_\odot$ to $\sim 10^{10}$M$_\odot$
enabling different science use cases and allowing for a number of
convergence tests.

The chosen cosmological parameters are: $\omega_{\rm cdm}=0.1109$,
$\omega_{\rm b}=0.02258$, $n_s=0.963$, $h=0.71$, $\sigma_8=0.8$,
$w=-1.0$. The volume of the simulation is $V=(4225.35~{\rm Mpc})^3=
(3000~h^{-1}{\rm Mpc})^3$, with 10,240$^3$=1.07 trillion
particles. This results in a particle mass of 
\begin{equation}
m_p=2.6\cdot 10^{9}~{\rm M}_\odot=1.85\cdot 10^9~h^{-1}{\rm M}_\odot.
\end{equation}
The size of the simulation was chosen to cover a volume large enough
to enable the generation of synthetic sky catalogs for surveys such as
DESI and LSST and at the same time to have enough mass resolution to capture
halos reliably down to small masses. The simulation was started at
$z_{in}=200$ and the initial conditions were generated using the
Zel'dovich approximation~\citep{Zel70}.  The transfer function was
generated with {\tt CAMB}~\citep{camb}. This leads to the same general
simulation set-up in all simulations mentioned above, Outer Rim,
Q Continuum, and MiraU.

\subsection{Particle Outputs -- Level 1 Data}

As for the Q Continuum and MiraU simulations, we store a large number
of time snapshots between $z=10$ and $z=0$. Originally, we saved 101
snapshots evenly spaced in $\log_{10}(a)$, but two snapshots were
unfortunately corrupted on disk -- an occupational hazard -- before
they were fully analyzed. The final list of 99 outputs in redshift $z$
is:
\begin{eqnarray}
z&=&\left\{10.04, 9.81, 9.56, 9.36, 9.15, 8.76, 8.57, 8.39, 8.05, 7.89, \right.\nonumber\\
&&7.74, 7.45, 7.31, 7.04, 6.91, 6.67, 6.56, 6.34, 6.13, 6.03, 5.84, \nonumber\\
&&5.66, 5.48, 5.32, 5.24, 5.09, 4.95, 4.74, 4.61, 4.49, 4.37, 4.26, \nonumber\\
&&4.10, 4.00, 3.86, 3.76, 3.63.  3.55, 3.43, 3.31, 3.21, 3.10, 3.04,\nonumber\\
&& 2.94, 2.85, 2.74, 2.65, 2.58, 2.48, 2.41, 2.32, 2.25, 2.17, 2.09, \nonumber\\
&&2.02, 1.95, 1.88, 1.80, 1.74, 1.68, 1.61, 1.54, 1.49, 1.43, 1.38, \nonumber\\
&&1.32, 1.26, 1.21, 1.15, 1.11, 1.06, 1.01, 0.96, 0.86, 0.82, \nonumber\\
&&0.78, 0.74, 0.69, 0.66, 0.62, 0.58, 0.54, 0.50, 0.47, 0.40, \nonumber\\
&&0.36, 0.33, 0.30, 0.27, 0.24, 0.21, 0.18, 0.15, 0.13, 0.10, 0.07, \nonumber\\
&&\left.0.05,0.02, 0.00\right\}.
\end{eqnarray}
For each particle in these snapshots we store position, velocity, and a unique 
particle ID to enable tracking particles over time. Each snapshot
encompasses roughly 40TB of data. 

Due to the wide variety of science projects enabled by a large
simulation, storing as much of the raw particle outputs as possible is
desirable. However, due to the large amount of data that is produced
it is not practical to keep the data for a long period of time on
spinning disk. In addition, data can get corrupted on disk. Therefore,
archiving data on long-term storage is important: For the Outer Rim
simulation, we use ALCF's and NERSC's tape storage system, which
employs the High-Performance Storage System (HPSS). Both facilities
have enabled Globus\footnote{https://www.globus.org/} on both disk and
tape storage system to allow easy transfer of the data. During the
period of running and analyzing the simulation, we lost two snapshots
on disk due to irreparable disk failures and also had several
corrupted failures on tape. A second copy on an independent tape
system therefore appears to be very important if one wants to
guarantee the availability of the full data set over an extended
period of time.

In addition to the full particle data sets, we also save for each time
snapshot a randomly selected set of particles that comprises 1\% of a
full snapshot. The down-sampled data sets are sufficient to measure
correlation functions and generate lightcone density maps for weak
lensing ray-tracing applications. The down-sampled data are kept on
disk for easy and fast access for post-processing analysis.

\subsubsection{Matter Power Spectrum }
\label{powspec}

\begin{figure}[t]
\centerline{
 \includegraphics[width=3.5in]{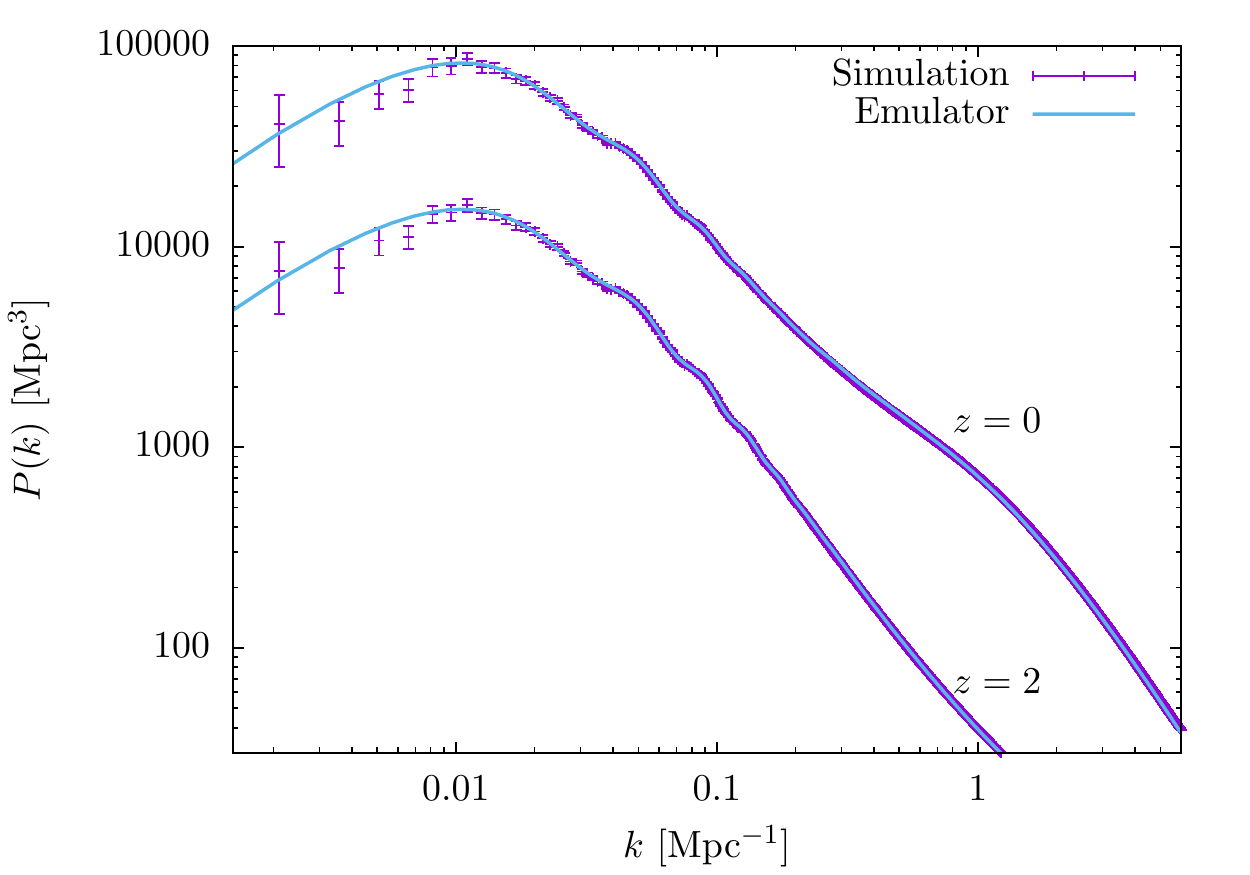}}
\caption{\label{fig:pow} Power spectrum results at redshift $z=0$ and
  $z=2$. For comparison, we show predictions from the Cosmic Emulator
  by~\cite{emu_ext}.} 
\end{figure}

The matter power spectrum was generated in-situ at many time snapshots
during the simulation. HACC is set up to automatically produce a power
spectrum before the end of a submission cycle and before every
check-point restart data dump. This allows for easy inspection of the
state of the simulation and whether the evolution is progressing as
expected.  In addition, for each time step at which we save a particle
snapshot we evaluate the matter power spectrum. Currently, the in-situ
power spectrum calculation is using the same grid size as the long-range
PM solver used in the simulation (in the current case, this is a
10,240$^3$ grid). Each power spectrum evaluation took approximately a
minute. Figure~\ref{fig:pow} shows the power spectrum for redshifts
$z=2$ and $z=0$ from the Outer Rim simulation. The large volume of the
simulation enables an excellent resolution for the baryonic
oscillation region (the ``wiggly'' region). We also show the
predictions from the Extended Cosmic Emulator, published
in~\cite{emu_ext}. As for our other simulations, Q Continuum and
Mira-Titan Universe, the agreement is very good and at the expected
level of accuracy following~\cite{emu_ext}.

\subsection{Friends-of-Friends Halo Outputs -- Level 2 Data}

Halo catalogs using different halo mass definitions were generated and
stored at each time snapshot in order to have sufficient information
to build detailed merger trees during the post-processing phase using
a new merger tree construction code~\citep{rangel17}. In the following we
provide details about the relevant data products.

\subsubsection{Friends-of-Friends Catalogs and Halo Particle Information}
\label{sec:fof}
The friends-of-friends (FOF) halo catalogs were generated using a
linking length of $b=0.168$. We store a large number of halo
properties -- halo position and velocity based on the halo minimum
potential as well as the center-of-mass, the halo mass, angular
momentum, and kinetic energy for all halos with at least 20
particles. The FOF finder follows the standard implementation of
identifying all particles that reside within a certain distance (the
linking length) of a particle and then the neighbors of those
particles, etc. The linking length is defined with respect to the mean
inter-particle spacing. This group finding algorithm was first used in
cosmology by~\cite{davis85}. The FOF algorithm is fast and makes no
assumptions about the halo geometry; it is also completely
reproducible -- two FOF finders run with the same linking length
should give exactly the same result. Our FOF halo finder
implementation~\citep{woodring11} is based on a very fast, tree-based serial halo
finder that is then parallelized by overloading each rank with a
sufficiently large border from the neighboring ranks to ensure that
halos that extend beyond one rank are successfully found. After all
halos are found on all ranks, halos that have been found more than
once (due to the overloading) are eliminated. In addition to the
``overloaded'' distribution of the particles, this is the only step in
the halo finding that requires communication between ranks.

The centers of the halos are determined by the location of the FOF
halo's minimum gravitational potential, where the potential at a given
halo particle is obtained by finding the distance $r$ to every other
halo particle and then accumulating all of the values of the negative of
mass divided by the distances to the particle. The center-of-mass and
the halo velocities are obtained by simply summing over all positions
and velocities and dividing by the number of particles. The FOF halo
mass is determined by the number count of particles in each
halo. Additional properties we store for each halo are the angular
momentum, ${\bf L}=\sum_i m_i ({\bf r}_i \times {\bf v}_i)$ and the
kinetic energy, $E=0.5\sum_i m_i {\bf v}_i^2$.  The implementation of
the halo finding algorithm and center finder are described in detail
(including timings for the very high mass resolution Q Continuum
simulation) in \cite{sewell15}. We follow a similar strategy for the
halo center finding as outlined in \cite{sewell15} for the last 14
time steps. We use 32 racks of Mira to find all the halos and
determine halo centers for halos up to 100,000 particles. For the more
massive halos, we write out the complete list of halo particles and
run the center finder on those outputs on only 8 racks. This reduces
the computational time considerably.

In addition to the halo catalogs themselves we also store all the
unique particle IDs and the associated halo id, if the particle
belongs to a halo. This information allows us to build halo merger
trees in post-processing explicitly following each particle's complete
history. We also save all the particles that reside in halos with at
least 100,000 particles (including position and velocity information
for all particles) and a random selection of 1\% of all halo
particles, but at least 5 particles per halo. This information is
useful when implementing HOD or similar galaxy models, allowing the
placement of model galaxies onto particles directly, if so desired.

Finally, we also store a handful of catalogs built with a linking
length of $b=0.2$. In particular, we generated catalogs at $z=\{$4.95,
4.0, 3.04, 2.02, 1.006, 0.0$\}$ with a minimum number of 20 particles
per halo. The main reason for analyzing halos with this specific
linking length is the existence of many results in the literature for
$b=0.2$, in order to compare our results with earlier
investigations. It also allows us to check the universality of the
mass function that was obtained in earlier work for this linking
length over a wide variety of redshifts. More details about the mass
function are given in Section~\ref{sec:massf}.

\subsubsection{Friends-of-Friends Halo Mass Function}
\label{sec:massf} 
 
\begin{figure}[t]
\centerline{\includegraphics[width=3.8in]{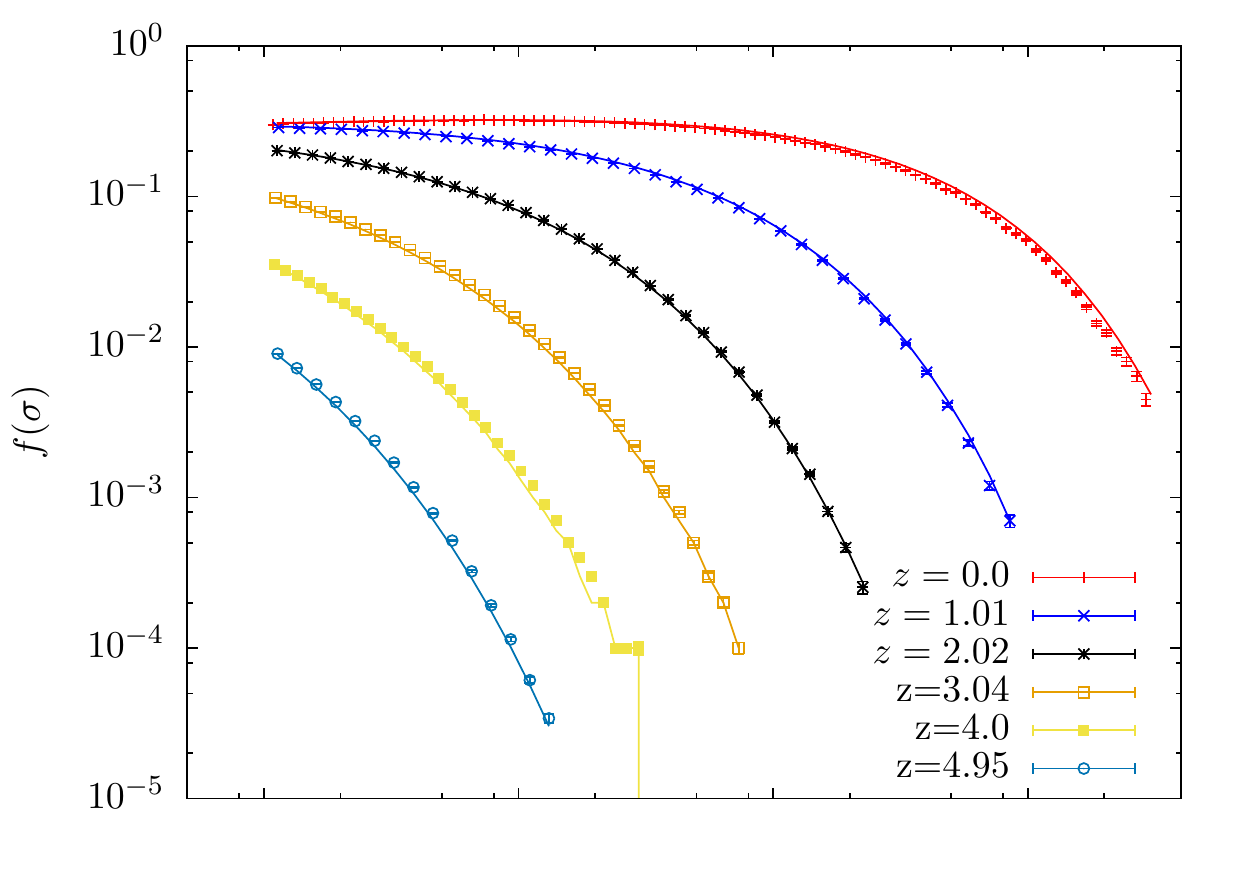}}
\vspace{-0.9cm}
\hspace{0.1cm}\centerline{\includegraphics[width=3.72in]{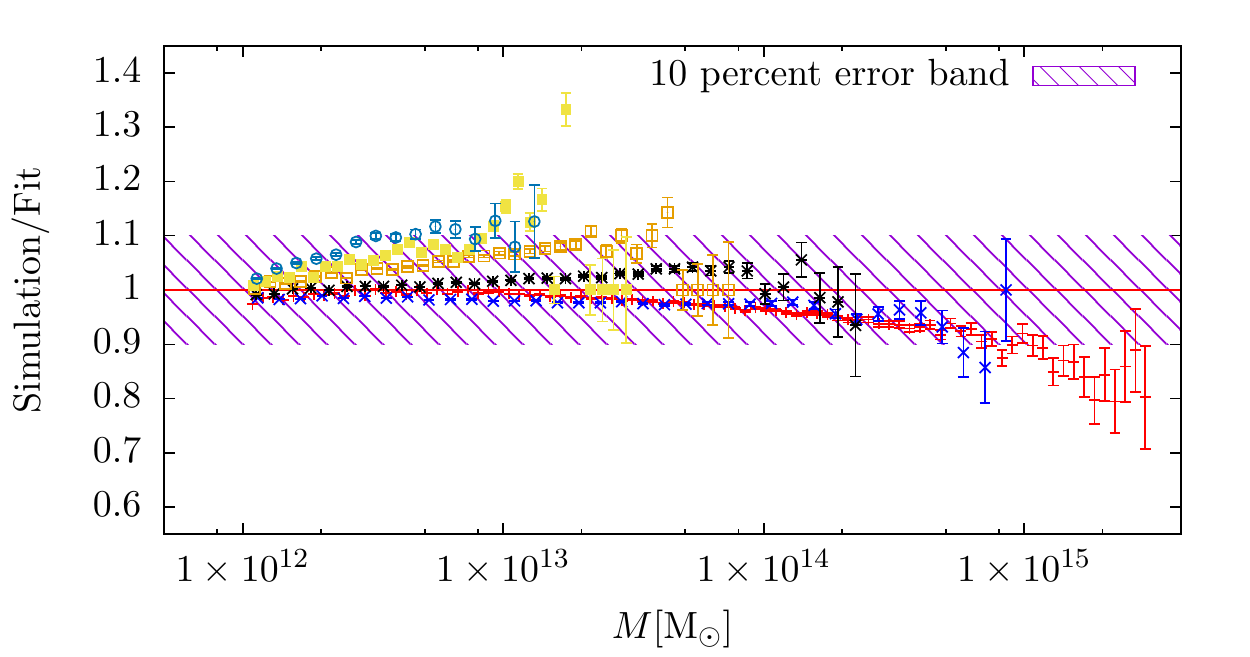}}
\caption{Upper image: Halo mass function $f(\sigma)$ as a function of
  halo mass at five redshifts. We show the measurements from the
  simulations (including statistical error bars) as well as the fit
  derived by \cite{bhattacharya11}. The mass definition used here is
  FOF with $b=0.2$. The lower image shows the ratio of the simulation
  result and the fit. We also mark a 10\% band that covers most of the
  results.\label{massf} }
\end{figure}

Next we show results for the halo mass function as measured by an FOF
halo finder with a linking length of $b=0.2$ to compare with a number
of previous results obtained using this definition and because of the
near-universality of the mass function obtained using this definition,
as first discussed in depth by~\cite{J01}. Figure~\ref{massf} shows
the differential mass function $f(\sigma,z)$ as introduced
by~\cite{J01} as a function of halo mass $M$ for redshifts between
$z=4.95$ and $z=0$: 
\begin{equation}
f(\sigma,z)=\frac{d\rho/\rho_b}{d\ln
  \sigma^{-1}}=\frac{M}{\rho_b(z)}\frac{dn(M,z)}{d\ln[\sigma^{-1}(M,z)]}. 
\end{equation}
Here, $n(M,z)$ denotes the number density of halos with mass $M$,
$\rho_b(z)$ is the background density at redshift $z$, and $\sigma
(M,z)$ is the variance of the linear density field.

\begin{figure}[t]
{\includegraphics[width=3.8in]{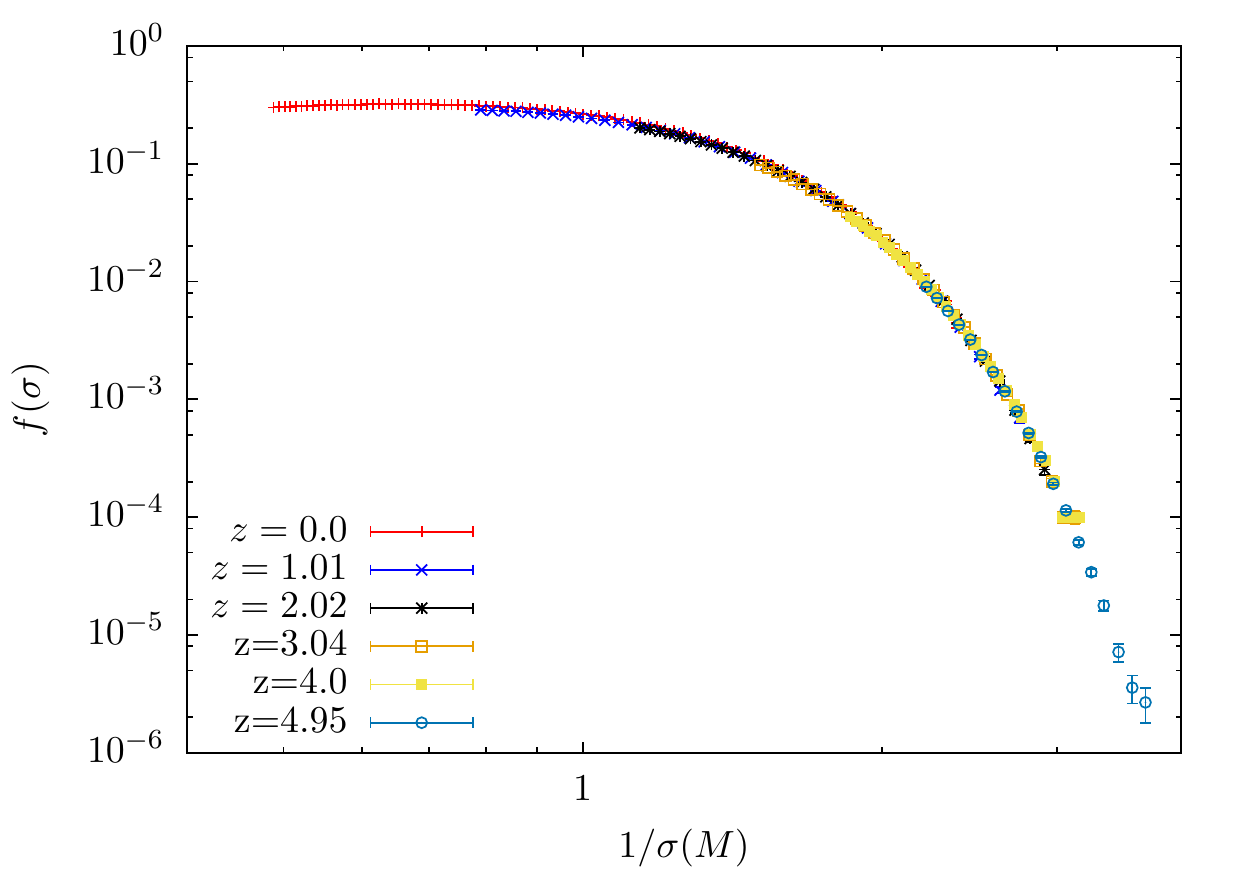}}
\caption{ Universality of the mass function across redshifts out to
  $z\sim 5$. We follow the behavior of the mass function $f(\sigma)$
  as a function of $1/\sigma(M)$ across a wide range of redshifts to
  demonstrate the remarkable power of universality. \label{fsigma}}
\end{figure}
The agreement across a wide range of redshifts over a broad dynamic
range (with the very small statistical errors) testifies to the power
of the universal description.

As is now commonly done for FOF halos (see,
e.g.,~\cite{bhattacharya11} for detailed tests) we include a mass
correction for low mass halos to account for finite sampling bias of
the form: 
\begin{equation}
n_h^{\rm corr} = n_h(1-n_h^{-0.65}).
\end{equation}
We only include bins with at least 100 halos. Besides the simulation
results, we also present a mass function fit derived
in~\cite{bhattacharya11}, including a simplification for the redshift
dependence suggested in~\cite{QCont}. The fit is given by: 
\begin{equation}\label{fit1}
f(\sigma,z)=A\sqrt{\frac{2}{\pi}}
\exp\left[-\frac{a\delta_c^2}
  {2\sigma^2}\right]\left[1+\left(\frac{\sigma^2}{a\delta_c^2}
  \right)^p\right] \left(\frac{\delta_c\sqrt{a}}{\sigma}\right)^q,
\end{equation}
with the parameters:
\begin{equation}\label{fit2}
A=\frac{0.333}{(1+z)^{0.11}};~~a=0.788;~~p=0.807,~~q=1.795.
\end{equation}
The density threshold for spherical collapse, $\delta_c=1.686$, is
taken to be the same for all redshifts.

The symbols in the upper panel in Fig.~\ref{massf} show the
measurements from the simulation, including statistical error bars,
while the lines show the mass function fit as given in
Eqs.~(\ref{fit1}, \ref{fit2}).  The lower panel in Fig.~\ref{massf}
shows the ratio of the simulation results and the mass function
fit. The shaded region marks a deviation of 10\% between fit and
simulation. In the cluster mass regime at $z=0$, the measured mass
function deviates from the fit at the 10\% level. This deviation is
consistent with earlier findings by, e.g., \cite{crocce10}, who
carefully compared different box size simulations and find differences
in the mass function for a 3$^{-1}$Gpc box compared to 4.5$^{-1}$Gpc
and 7.68$^{-1}$Gpc boxes at a similar level as we find here.  In order
to demonstrate the almost universal form of the mass function more
concretely, we also plot $f(\sigma,z)$ as a function of $1/\sigma(M)$
in Fig.~\ref{fsigma} for all six redshifts between $z=4.95$ and $z=0$.

\subsubsection{Halo Core Catalogs}

\begin{figure}
\centerline{
 \includegraphics[width=3.5in]{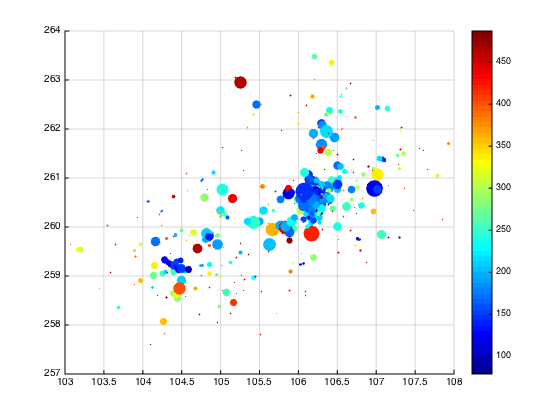}}
\caption{\label{cores}Halo cores found in a cluster sized halo with a
  mass of approximately $10^{15}$M$_\odot$ at redhift $z=0$. The core
  of each halo with more than 100 particles that fell into this
  cluster at some point in time is marked with a point. The point size
  is 100 times the core radius and the color indicates the redshift of
  infall. As expected from hierarchical structure formation, the
  center of the cluster is populated with cores from very early times
  while the outskirts are mostly populated with halo cores that more
  recently fell into the cluster.}
\end{figure}

The halo formation history plays a crucial role in influencing the
properties of the galaxies hosted by a particular halo~(see, e.g.,
\cite{wechsler18} for a recent review). In particular, the infall mass
of a halo when joining another halo and becoming a part of it, either
in the form of a subhalo or simply in the form of additional,
disrupted dark matter mass, is important to follow over time. In order
to fully keep track of this information, including the afterlife of a
halo once it has fallen into another halo, we introduce the notion of
a halo ``core''. For each halo that is larger than 100 particles at a
time step $i$, we determine the position and velocity of the ten
particles closest to the center, the halo core particles. We store a
halo core file for each time snapshot reporting all halo core
particles, including unique particle IDs. In addition, at each of the
remaining time steps, we identify the positions of all the core
particles from all the previous time snapshots and store them in an
accumulated core particle file. If a particle is still a core particle
in step $i+j$ (for example the halo at step $i$ never merged or fell
into another halo but rather just accreted mass) we only record the
core particles once in the accumulated core particle file. The
accumulated core particle files allow us to track halos over time in
great detail. The advantage of the core tracking over simple subhalo
merger trees is that we retain the information of disrupted halos as
well. We carried out detailed tests on a smaller simulation with
similar mass resolution to confirm that ten core particles provide
this information reliably. We will present a detailed study of the
core particle tracking and comparison to subhalo identification in a
forthcoming paper.

\begin{figure}[t]
\centerline{
 \includegraphics[width=3.5in]{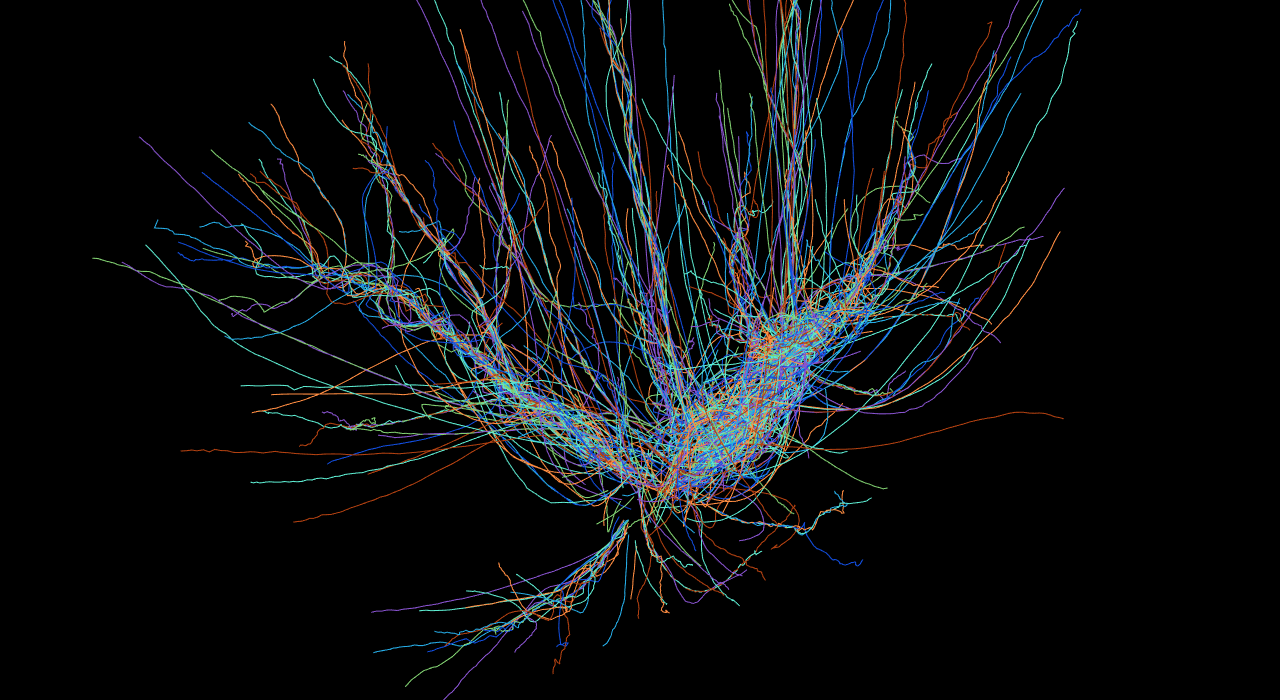}}
\caption{\label{fig:traj} Halo core trajectories followed over time
  leading to one massive halo at $z=0$. Each line tracks the evolution
  of a core from a halo that had at least 100 particles at formation
  time until its infall into the main halo. }
\end{figure}

Figure~\ref{cores} shows an example of the cores identified in a
cluster sized halo of mass~$\sim 10^{15}$M$_\odot$. Each point shows a
core that originated at some point from a 100 particle halo. The cores
are colored with respect to their infall time while the size signals
the radius of the core (magnified by a factor of 100 for
visibility). The cores close to the center belong the longest to the
mother halo as one would expect. Most halos on the outskirts joined
the main halo at later times. Figure~\ref{fig:traj} shows the
trajectories of cores over time joining to merge in one large halo at
the end. Each core trajectory follows a halo of at least 100 particles
until it falls into another halo. 

\subsubsection{Subhalo Catalogs}

In addition to the halo catalogs, we have also identified subhalos for
cluster sized halos with more than 100,000 particles for redshifts
between $1.5\le z\le 0$. The redshift range and mass cut are chosen so
as to model clusters as observed by the South Pole Telescope (SPT) and
described in the SPT cluster paper by~\cite{bleem15}. We find subhalos
by combining a local density estimator with phase space
information. After building a list of possible subhalo candidates we
determine if a particle is gravitationally bound to the subhalo; if it
does not belong to the subhalo, it is assigned to the main halo. We
then keep only subhalos that have at least 20 particles. We have used
this approach in~\cite{li15} to find subhalos and then populated three
clusters from the Outer Rim simulation with cluster galaxies. We
generated realistic strong lensing images from these clusters
emulating the known properties of different telescopes. With the
complete set of clusters, including subhalo information, we will
extend the work described in~\cite{li15} in the near future to a
complete SPT cluster sample.  In addition, the subhalo information
will be used for comparison with the core tracking approach described
in the previous section. Preliminary results show that intact cores
within major halos can be mapped onto subhalos found by the subhalo
finder. A detailed analysis of these results is in preparation.

\subsubsection{Merger Trees}

In addition to the halo core tracking, we also built traditional
merger trees on the FOF halo catalogs. We take into account each halo
with at least 20 particles and follow its evolution over time. For
each halo at $z=0$ we have a complete merger history as far back into
the past as $z=10$. As is well known, building halo merger trees on a
number of discrete time snapshots is very challenging. Halos fall in
and out of existence due to an unavoidable lower mass bound, there are
halo merging and splitting events, and fly-by events where parts of
halos may get disrupted. It is therefore important to build merger
trees taking into account not only two neighboring time snapshots but
by following the evolution of the halos over several time steps. In
addition, the enormous amount of data that has to be processed for a
simulation like the Outer Rim simulation requires careful
implementation of the merger tree algorithm including an optimized
load-balancing scheme. Our methodology, and initial results, are
described in~\cite{rangel17}; a longer paper is in preparation.

\subsection{Overdensity Catalogs -- Level 2 Data}

In addition to the FOF catalogs, we also store halo catalogs with
overdensity masses. For each of the 99 snapshots we measure
overdensity masses M$_{200c}$, defined with respect to 200 times the
critical density. The halo location is based on FOF halo centers
(determined from the potential minimum center, as described above) and
SO masses are determined for halos with at least 1,000 particles. (The
SO halo itself is measured from the full particle output, not just
from the FOF particles.) We store a range of halo properties including
the radius, the mass, the kinetic energy and angular momentum, as
described previously (but now calculated from the SO halo particles),
the halo concentration, the halo profile, and the velocity
dispersion. For all snapshots out to $z\sim 1.5$ we also measure
M$_{500c}$, again with a view of generating a cluster catalog to match
the SPT survey. 

\subsubsection{Concentration-Mass Relation}

We have measured the concentration of all M$_{200c}$ halos with more
than 1000 particles; different concentration measurement approaches
and results over a wide range of redshifts are discussed
in~\cite{child18}, where we also carry out a detailed analysis of
possible systematics due to the set up of the simulation (e.g.,
starting redshift, time stepping, halo centering) and use results from
a range of simulations, including the Q Continuum and the MiraU
simulations. \cite{child18} provides comparison with observational
results as well as with fitting functions from a range of previous
simulation based studies.

\subsection{Lightcone Catalogs and Maps -- Level 3 Data}

\begin{figure}[t]
\centerline{
 \includegraphics[width=3.5in]{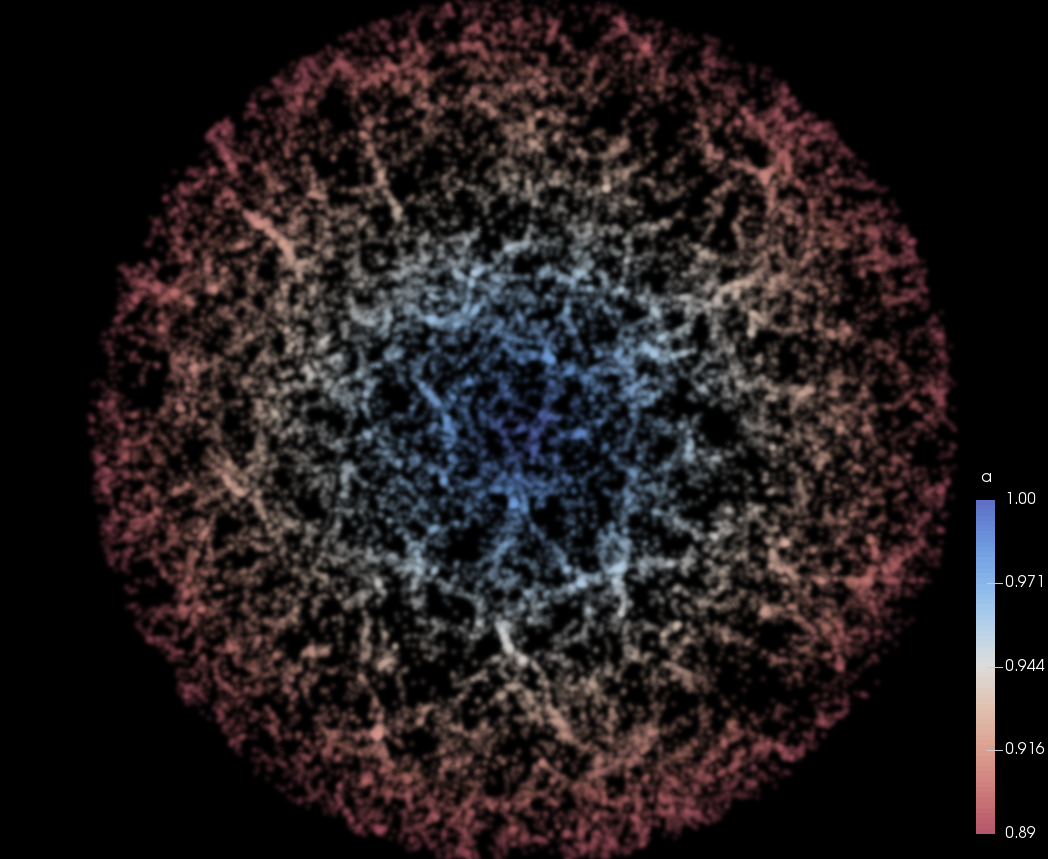}}
\caption{Visualization of the M$_{500c}$ lightcone catalog out to
  redshift $z\sim 0.13$ and downsampled to 6 percent of the data to
  enhance the visibility of the cosmic web. The data is colored with
  respect to the value of the scale factor $a$ for each
  halo.\label{fig:lightc}}
\end{figure}

The large range of Level~1 and Level~2 data allows us to generate
synthetic sky maps at different wavelengths. Figure~\ref{fig:lightc}
shows the visualization of a subset of the lightcone data from our
M$_{500c}$ catalog. We show the data out to a moderate redshift of
$z\sim0.13$ (close to what the Sloan Digital Sky Survey main sample
covered) and reduce the number of halos down to 6~percent of the data
from the full sphere to visually emphasize the structure of the cosmic
web. The full dataset is currently used for a major strong lensing
project including data from the South Pole Telescope and follow-up
optical observations. For a recent paper on generating strong lensing
images from the Outer Rim simulation, see \cite{li15}, and for the
investigation of the influence of line of sight halos on the
detectability of cluster-scale strong lensing, see \cite{li18}. Based
on this work, a large set of strong lensing images has been generated
and predictions for the number of strong lenses that should be seen by
SPT and follow-up optical observations are being derived and compared
to the actual observations.

We have also recently generated optical catalogs using a new approach
that combines empirical and semi-analytic methods. The new method
makes highly efficient use of compute resources by leveraging results
from smaller, high-resolution simulations into very large volume,
high-resolution simulations like the Outer Rim simulation. The concept
is rather straightforward, the details of the implementation will be
described in a forthcoming paper. The empirical approach is used to
populate the halos in the Outer Rim simulation with galaxies to obtain
the correct clustering statistics and to assign basic properties such
as colors to match to observational data. Then the resulting galaxies
are matched up with galaxies from a comprehensive catalog of galaxies
generated with a semi-analytic approach to create a final catalog with
a rich set of galaxy properties.  For the semi-analytic part, we use
Galacticus, developed by~\cite{benson}. Galacticus follows the
evolution of each halo over time (in post-processing) and by applying
a range of well-motivated models mimics the astrophysical processes
that will lead to the formation and evolution of galaxies. We ran
Galacticus on a downscaled version of the Outer Rim simulation (a
(256$h^{-1}$Mpc)$^3$ volume with 1024$^3$ particles) and generated a
galaxy library this way out to $z=3$. The size of the Outer Rim
simulation allows us to generate a large optical catalog employing
this new approach.

The first use of the resulting optical catalog will be for the LSST
Dark Energy Science Collaboration data challenges.  These data
challenges are full end-to-end simulations of LSST catalogs, starting
from an N-body simulation, through image simulation tools, and finally
include processing with the LSST data management stack. The full
effort will be described in detail elsewhere. In addition, the
simulation has been used to generate mock catalogs for DESI for
testing fiber assignment algorithms and to optimize the DESI survey
strategy (for an image of the catalog data see \citealt{habib14}). The
catalog generation for this was based on a halo-occupancy distribution
(HOD) approach. The eBOSS collaboration has created mock catalogs for
their analysis using an HOD approach for a variety of projects. These
projects include the clustering measurement of quasars
by~\cite{eboss1} via the correlation function and by~\cite{eboss2} via
the power spectrum and by~\cite{eboss3} focusing on the anisotropic
clustering of the eBOSS quasar sample in configuration space.

\begin{figure}[t]
\centerline{
 \includegraphics[width=3.5in]{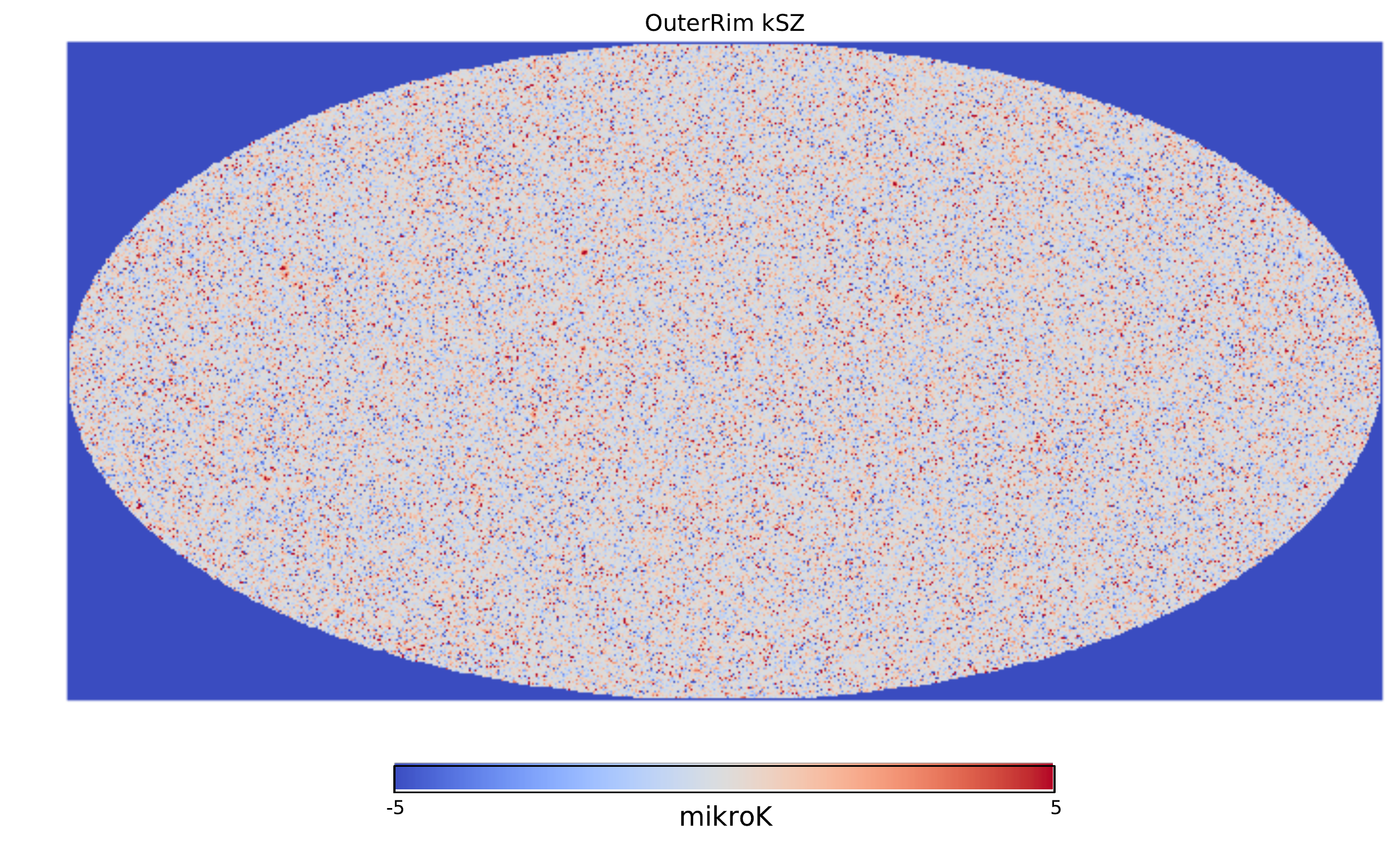}}
 \centerline{\includegraphics[width=3.5in]{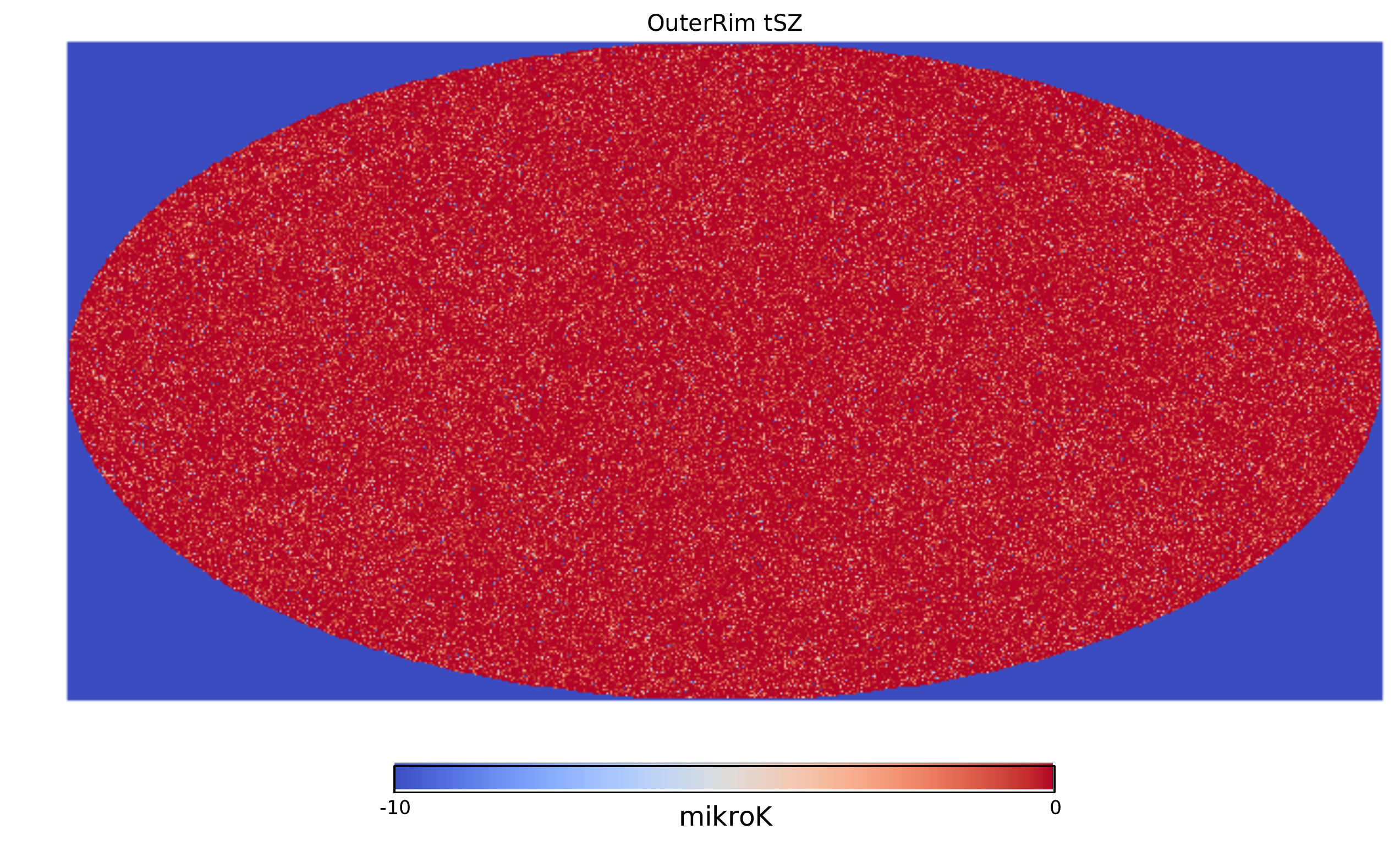}}
 \caption{Upper panel: full-sky map of the kinematic
   Sunyaev-Zel'dovich effect based on the Outer Rim simulation. Lower
   panel: Corresponding all-sky map of the thermal Sunyaev-Zel'dovich
   effect. \label{fig:SZ}}
\end{figure}

Beyond optical synthetic maps, the Outer Rim simulation also lends
itself to generation of maps for CMB observations. Figure~\ref{fig:SZ}
shows an example for maps of the kinematic and thermal
Sunyaev-Zel'dovich effect. The map making process is based on the
procedure described in detail in~\cite{flender}. The high mass
resolution and the large volume covered by the simulations also
enables us to model cosmic infrared background sources, a task that
has been difficult to achieve in the past due to lack of resolution
and/or volume covered. These are just a few examples for the large
variety of projects that can be carried out with data from the Outer
Rim simulation. The data set will be very valuable in the future to
build catalogs for a range of surveys to help test pipelines and
validate cosmology analysis tools.

\section{Data Release}
\label{sec:release}

As part of this paper we release some of the data products publicly. A
more detailed description about the data products and how to read the
data is given in \cite{heitmann18}. The data release paper
\citep{heitmann18} includes several more simulations besides the Outer
Rim simulation. For the Outer Rim simulation we provide time snapshots
of down-sampled particle files and FOF halo catalogs for the following
redshifts:
\begin{equation}
z=\{0.0, 0.05,0.21,0.50,0.78,0.86,1.43\}.
\end{equation}
We also provide particle lightcone and halo lightcone data for one
octant of a sphere.  The data is stored in GenericIO format, a reader
is provided that can be used to read the data into a python code or to
convert the files to ASCII (the user is warned that for the
downsampled particles this might not be a good idea). The downsampled
particle data files are 0.4TB each, the halo catalogs vary with
redshift. Each output includes metadata, describing the content of the
files in detail. The halo catalogs contain information about the
number of particles in each halo, a halo ID (which is by itself not
meaningful), the halo mass measured in $h^{-1}$M$_\odot$, the halo
center given by the potential minimum, the center of mass coordinates
and velocities, and the halo velocity dispersion. Centers are measured
in units of $h^{-1}$Mpc and velocities are given in comoving peculiar
velocities measured in km/s. The particle files contain particle
positions, velocities, IDs, a potential value, and a mask. The
potential value is measured on the PM grid that was used for the
simulation and has not been normalized.

In order to access these data products, please visit our webpage at
https://cosmology.alcf.anl.gov/, choose the Outer Rim simulation and
follow the instructions given there and in~\cite{heitmann18}.

\section{Conclusions and Outlook}
\label{conclusion}

In this paper we introduced the Outer Rim simulation, one of the
largest cosmological simulations currently available at the volume and
mass resolution covered. The simulation generated more than 5PB of raw
data and we have extracted a range of useful data from the raw
particle outputs for further science projects (Level~2 and Level~3 in
the parlance of the paper). The detailed information about halos,
their evolution over time as captured by merger trees, and the fate of
their inner cores allows for the construction of sophisticated
synthetic sky maps in different wavebands. We have showcased some
examples here such as maps of the Sunyaev-Zel'dovich effect, but many
more catalogs are currently being constructed by several science
collaborations.

The simulation was carried out on Mira, a BG/Q system at the Argonne
Leadership Computing Facility. In order to enable such a large
simulation, our cosmology code HACC had to optimized in many ways,
ranging from optimal memory usage, to fast I/O, to customization of
the force kernel for the BG/Q architecture, to name just a few. We
have provided some information of these optimizations in this paper
with a view to proceeding to next generation machines.

Running a large simulation on basically the full machine was a major
challenge, as was the analysis of the resulting data. We have carried
out some of the analysis in-situ to minimize I/O times and some of the
analysis in post-processing. When carrying out in-situ analysis it is
very important that the tools perform at very high efficiency, forcing
the development of load-balanced approaches that enable the most
efficient use of the full machine. As part of this paper, we make some
of the data products from the simulation publicly available. With
time, we will make many of the synthetic catalogs based upon the
simulation available as well.

\begin{acknowledgments}

Argonne National Laboratory's work was supported under the
U.S. Department of Energy contract DE-AC02-06CH11357. N. Frontiere
acknowledges support from the DOE CSGF Fellowship program.  This
research used resources of the Argonne Leadership Computing Facility
at the Argonne National Laboratory, which is supported by the Office
of Science of the U.S. Department of Energy under Contract
No. DE-AC02-06CH11357. We are indebted to the ALCF team for their
outstanding support and help to enable us to carry out a simulation at
this scale. The research described in the paper was partially
supported by the Scientific Discovery through Advanced Computing
(SciDAC) program funded by the U.S. DOE Office of Advanced Scientific
Computing Research and the Office of High Energy Physics. We thank Jon
Woodring for visualization support.

\end{acknowledgments}

\end{document}